\documentclass[12pt,preprint]{aastex}

\newcommand{\element}[2]{$^{#2}\mathrm{#1}$}

\shorttitle{NLTE boron abundances in metal-poor stars}
\shortauthors{Tan, Shi, \& Zhao}

\begin{document}

\title{A NLTE analysis of boron abundances in metal-poor stars\altaffilmark{\star}}

\author{Kefeng Tan\altaffilmark{1,2}, Jianrong Shi\altaffilmark{1},
    and Gang Zhao\altaffilmark{1}}

\altaffiltext{$\star$}{Based on observations made with ESO telescopes
    and NASA/ESA {\it Hubble Space Telescope} obtained from the ESO/ST-ECF
    Science Archive Facility; based on spectral data retrieved from
    the ELODIE archive at Observatoire de Haute-Provence (OHP).}
\altaffiltext{1}{Key Laboratory of Optical Astronomy, National
    Astronomical Observatories, Chinese Academy of Sciences, Beijing
    100012, China; [tan, sjr, gzhao]@bao.ac.cn}
\altaffiltext{2}{Graduate University of Chinese Academy of Sciences,
    Beijing 100049, China}

\begin{abstract}
The non-local thermodynamic equilibrium (NLTE) line formation of
neutral boron in the atmospheres of cool stars are investigated. Our
results confirm that NLTE effects for the \ion{B}{1} resonance lines,
which are due to a combination of overionization and optical pumping
effects, are most important for hot, metal-poor, and low-gravity stars;
however, the amplitude of departures from LTE found by this work are
smaller than that of previous studies. In addition, our calculation
shows that the line formation of \ion{B}{1} will get closer to LTE if
the strength of collisions with neutral hydrogen increases, which is
contrary to the result of previous studies. The NLTE line formation
results are applied to the determination of boron abundances for a
sample of 16 metal-poor stars with the method of spectrum synthesis of
the \ion{B}{1} 2497\,{\AA} resonance lines using the archived
{\it HST}/GHRS spectra. Beryllium and oxygen abundances are also
determined for these stars with the published equivalent widths of the
\ion{Be}{2} 3131\,{\AA} resonance and \ion{O}{1} 7774\,{\AA} triplet
lines, respectively. The abundances of the nine stars which are not
depleted in Be or B show that, no matter the strength of collisions
with neutral hydrogen may be, both Be and B increase with O
quasi-linearly in the logarithmic plane, which confirms the
conclusions that Be and B are mainly produced by primary process in
the early Galaxy. The most noteworthy result of this work is that B
increases with Fe or O at a very similar speed as, or a bit faster
than Be does, which is in accord with the theoretical models. The B/Be
ratios remain almost constant over the metallicity range investigated
here. Our average B/Be ratio falls in the interval $[13\pm4, 17\pm4]$,
which is consistent with the predictions of spallation process. The
contribution of B from the $\nu$-process may be required if the
\element{B}{11}/\element{B}{10} isotopic ratios in metal-poor stars
are the same as the meteoric value. An accurate measurement of the
\element{B}{11}/\element{B}{10} ratios in metal-poor stars is crucial
to understanding the production history of boron.
\end{abstract}

\keywords{Galaxy: evolution --- line: formation --- line: profiles ---
    stars: abundances --- stars: atmospheres --- stars: Population~II}

\section{Introduction}\label{intro}

The origin and evolution of boron (and beryllium) are of special
interest because they are hardly produced by the standard big bang
nucleosynthesis (BBN), nor can they be produced by nuclear fusions in
stellar interiors. \citet{reeves1970} were the first to propose that
most of the light elements (Li, Be, and B) can be produced by the high
energy processes involving Galactic cosmic rays (GCR) acting on the
interstellar medium (ISM). Subsequent detailed model by
\citet{meneguzzi1971} could reproduce most of the observations at that
time; one of the exceptions is that the theoretical isotopic abundance
ratio \element{B}{11}/\element{B}{10}\footnote{$\mathrm{A/B}=N\mathrm
{(A)}/N\mathrm{(B)}$} (2.4) is lower than the observed
solar meteoritic value of about 4.0. In order to solve this problem,
they introduced a non-observable intense low energy component to the
GCR. \citet{woosley1990} proposed that \element{Li}{7} and \element{B}{11}
can be synthesized by neutrino spallation in the helium and carbon
shells of supernovae, i.e., the so-called $\nu$-process, which argues
against the existence of an unobserved low energy component of GCR.
Nevertheless, the GCR process, which predicts a quadratic relation
between B (also Be) and O abundances, remained as the standard picture
for light elements production until the late 1980's. However, later
observations on B \citep{duncan1992,edvardsson1994,duncan1997,
duncan1998b,garcialopez1998}, as well as on Be in metal-poor stars
showed that B and Be abundances increase quasi-linearly with O
abundances, which indicates that B and Be are probably produced by
primary process, rather than the standard secondary GCR process.

All the studies on B abundances mentioned above were based on the
ultraviolet (UV) \ion{B}{1} resonance lines at 2497\,{\AA}.
\citet[hereafter K94]{kiselman1994} investigated the formation of the
\ion{B}{1} resonance lines in the atmospheres of solar-type stars and
found significant departures from LTE in stars hotter or more
metal-deficient than the Sun. In their subsequent work,
\citet[hereafter KC96]{kiselman1996} presented the NLTE abundance
corrections for the \ion{B}{1} 2497\,{\AA} resonance lines for a grid
of cool stellar model atmospheres. After that, in almost all the
studies on boron, B abundances were determined in LTE first and then
corrected with the results of KC96. However, NLTE corrections are
model dependent and in principle should be applied only to the
analysis using the same model atmospheres. Moreover,
\citet{boesgaard2004} pointed that B abundance with the NLTE
corrections of KC96 increases more slowly for halo stars than the Be
abundance does, which is not predicted by light-element synthesis or
depletion, therefore they suggested a full NLTE analysis for B, rather
than simply applying corrections to the LTE abundances.

The aim of this work is thus to re-investigate the NLTE line formation
of neutral boron in the atmospheres of cool stars first, and then to
obtain the NLTE B abundances for a sample of metal-poor stars, based
on which the origin and evolution of B in the early Galaxy can be
investigated. In $\S$~2 we briefly describe the observations and data
reduction. Section~\ref{atmos} presents the adopted model atmospheres
and stellar parameters. Section~\ref{nlte} deals with the NLTE line
formation of neutral boron. Section~\ref{results} gives the abundances
and uncertainties. In $\S$~\ref{discuss} we discuss the implications
for the origin and evolution of boron, while in the last section we
briefly summarize our results and conclusions.

\section{Observations and data reduction}\label{obs}

Our analysis of B abundances are based on the archived spectra of 16
cool metal-poor stars ($-2.7<\mathrm{[Fe/H]}\footnote{$\mathrm{[A/B]}
=\log[N(\mathrm{A})/N(\mathrm{B})]_{\bigstar}-\log[N(\mathrm{A})/
N(\mathrm{B})]_{\odot}$}<-0.5$), which were obtained with the Goddard
High Resolution Spectrograph (GHRS) on board the {\it Hubble Space
Telescope} ({\it HST}). The sample stars were selected from four
original studies on boron abundances \citep{duncan1992, duncan1997,
garcialopez1998,primas1999}. All the stars were observed using the
G270M grating centered at the \ion{B}{1} 2497\,{\AA} resonance doublet
with a resolution of about 25,000. The signal-to-noise ratios (S/N) of
the spectra are higher than 50 per diode. More details about the
observations can be retrieved from the {\it HST} Science Archive
Database and the aforementioned original paper. The spectra were
reduced following the standard {\it HST} procedure using the STSDAS
package of the IRAF\footnote{IRAF is distributed by the National
Optical Astronomy Observatories, which are operated by the Association
of Universities for Research in Astronomy, Inc., under cooperative
agreement with the National Science Foundation.} suite of programs.

\section{Model atmospheres and stellar parameters}\label{atmos}

We adopted the opacity sampling (OS) model atmosphere MAFAGS-OS9
presented by \citet{grupp2009} both in the NLTE line formation
investigation and in the abundance analysis. This model is based on
the one-dimensional (1D) LTE theoretical model atmosphere MAFAGS-OS
described by \citet{grupp2004} but incorporates the \citet{kurucz2009}
iron atomic data as well as the new bound-free ($b$-$f$) opacities for
\ion{Mg}{1} and \ion{Al}{1}. The emergent solar flux from this new
model shows significantly better agreement with observations,
especially in the UV region of the spectrum, as can be seen in
Figure~3 of \citet{grupp2009}.

Stellar parameters are fundamental to deriving elemental abundances in
the photospheres of stars. However, it's impractical to determine them
accurately with the narrow-wavelength-range (about 40\,{\AA}) and
severely-blended UV {\it HST}/GHRS spectra. We noted that 11 of our sample
stars had been studied previously by \citet{korn2003},
\citet{gehren2006}, \citet{mashonkina2008}, and \citet{tan2009}, where
the stellar parameters were determined spectroscopically using almost
the same model atmosphere code and technique, i.e., effective
temperatures were derived by fitting the wings of Balmer lines;
surface gravities were based on the {\it Hipparcos} parallaxes
\citep{hipparcos} if available\footnote{Only one star BD\,$-13\degr 3442$
has no {\it Hipparcos} parallax data; its surface gravity was
determined from the ionization balance between Ca\,{\sc i} and
Ca\,{\sc ii} by \citet{mashonkina2008}.}; iron abundances were determined
from \ion{Fe}{2} lines; and the microturbulent velocities were estimated
by requiring the iron abundances independent of equivalent widths or by
fitting the strongest \ion{Fe}{2}, \ion{Ca}{1}, and \ion{Mg}{1} lines
which are sensitive to microturbulent velocities. For the rest five
stars, fortunately, high-resolution and high S/N spectra covering a
wide wavelength range were available from the archived ESO VLT/UVES
\citep{uves} or OHP ELODIE \citep{elodie} spectra
database. In order to keep our analysis consistent, we determined the
stellar parameters of these five stars using the method described
above. The final stellar parameters are given in Table~\ref{table:par}.
As discussed in \citet{tan2009}, the typical uncertainties for the
effective temperature, surface gravity, iron abundance, and
microturbulent velocity are estimated to be $\pm$80\,K, $\pm$0.15\,dex,
$\pm$0.08\,dex, and $\pm$0.2\,km\,s$^{-1}$ respectively.

\section{NLTE line formation of neutral boron}\label{nlte}

\subsection{Boron atomic model}\label{atom}

\begin{figure}
\centering
\includegraphics{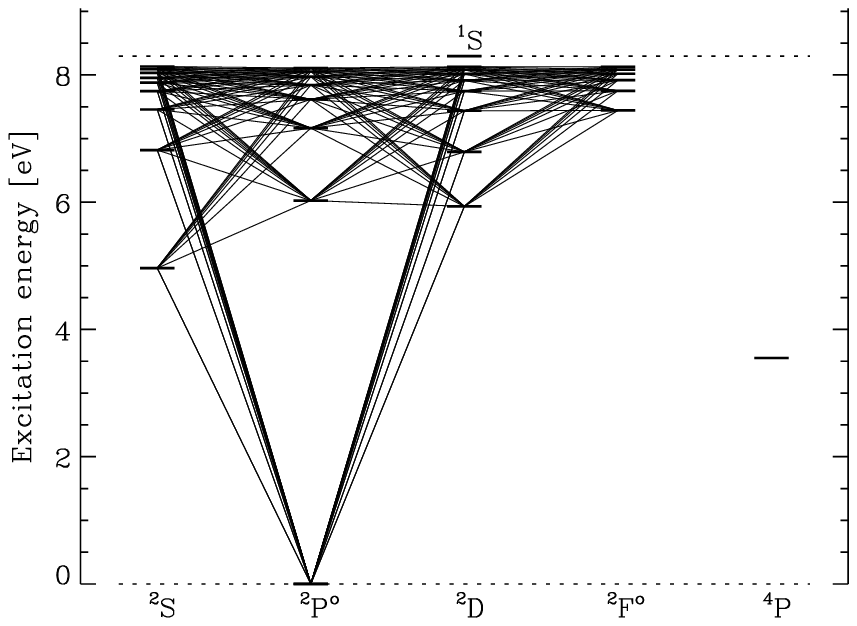}
\caption{Grotrian diagram for \ion{B}{1} (the ground state of
\ion{B}{2} is also shown). The continuous lines denote the
radiative bound-bound transitions considered in this work.
\label{fig:grotrian}}
\end{figure}

The boron atomic model adopted in this work is illustrated by its
Grotrian diagram as shown in Figure~\ref{fig:grotrian}. Details for
each item, as well as a comparison with the counterpart of K94 are
described in the following paragraphs.

\subsubsection{Energy levels}

The energy levels were taken from the laboratory data compiled by
\citet{kramida2007}. Fine structure splitting was neglected except for
the ground state 2p~$^2\mathrm{P^o}$. The final model atom includes 33
\ion{B}{1} bound levels plus the \ion{B}{2} ground state as the
continuum. In fact, for testing purpose, we have extended our model by
adding 61 \ion{B}{2} bound levels and the \ion{B}{3} ground state as
the continuum, but the calculated results showed that there are no
essential differences. The final boron atomic model adopted by K94
includes 30 bound levels and one continuum level, which is comparable
to that of this work. The energy levels of K94 were taken from the
laboratory data compilation of \citet{bashkin1978} except for the
higher $^2\mathrm{P^o}$ levels which were taken from the TOPbase
\citep{topbase}. The mixing of laboratory and theoretical data makes
the energy level sequence of K94 a bit different from that of this
work, for example, their 7s~$^2\mathrm{S}$ level is higher than
7p~$^2\mathrm{P^o}$, and the 8s~$^2\mathrm{S}$ level is higher than
8p~$^2\mathrm{P^o}$, which are contrary to that of \citet{kramida2007}.

\subsubsection{Radiative transitions}

Oscillator strengths were extracted from the TOPbase except for the
transitions 2p~$^2\mathrm{P^o}$--3s~$^2\mathrm{S}$ and
2p~$^2\mathrm{P^o}$--2p$^2$~$^2\mathrm{D}$, which have accurate
laboratory data from \citet{johansson1993}. K94 adopted exactly the
same data as ours except that they excluded the lines with $f<0.001$
and the lines of extremely long wavelength.

Photoionization cross-sections were available from the TOPbase for 26
levels; for the rest levels (the $^4\mathrm{P}$ and $^2\mathrm{F^o}$
levels) hydrogenic approximations were adopted. K94 also adopted the
photoionization cross-sections from the TOPbase when available, but
their data for the $^4\mathrm{P}$ level was provided by
Dr.~K.~Berrington and the $^2\mathrm{F^o}$ levels' cross-sections were
copied directly from the $^2\mathrm{P^o}$ levels with similar energies.

\subsubsection{Collisional transitions}

For electron impact excitation, the effective collision strengths were
available from the $R$-matrix with pseudostates (RMPS) calculations of
\citet{ballance2007} for the transitions among the first 12 levels;
the remaining allowed and forbidden transitions were approximated by
the formulae of \citet{vanregemorter1962} and \citet{allen1973},
respectively. Electron impact ionization cross-sections were
calculated from the classical path approximation of \citet{seaton1962}.
K94 adopted the electron impact excitation cross-sections calculated
by \citet{nakazaki1991} for the transitions among the first nine
levels; the approximation of \citet{vanregemorter1962} was adopted for
the remaining transitions. The electron impact ionization rates of K94
were calculated from the hydrogenic approximations of \citet{allen1973}.

Excitation and ionization induced by inelastic collisions with neutral
hydrogen atoms are usually calculated using the formula of
\citet{drawin1968,drawin1969} presented in the form for astrophysical
applications by \citet{steenbock1984}. However, it has been shown by
\citet[for \ion{Na}{1}]{belyaev1999} and \citet[for \ion{Li}{1}]{belyaev2003}
that collision rates of the resonance transitions given by Drawin's
formula are overestimated by orders of magnitude. Therefore the
collision efficiency was commonly constrained by applying a scaling
factor $S_{\mathrm{H}}$ to the Drawin's formula empirically, and the
factor could be determined by requiring the derived abundances from
different lines to be consistent \citep{zhao1998,zhao2000}. In this
work, the cross-sections are computed with three different scaling
factors, namely $S_{\mathrm{H}}=0$ (no collisions with neutral
hydrogen), 0.1, and 1, but as there are only two observed \ion{B}{1}
lines, and one is severely blended, we are not able to determine the
final $S_{\mathrm{H}}$ value using the method described above. However,
as will be discussed in $\S$~\ref{discuss}, the uncertainty in the
strength of collisions with neutral hydrogen does not affect our final
conclusions much. For K94, the cross-sections for the $b$-$f$
transitions induced by collisions with neutral hydrogen were
calculated according to the analytic expressions of \citet{kunc1991},
while the bound-bound ($b$-$b$) transitions were neglected.

\subsection{NLTE line formation results}\label{nlte_result}

The coupled radiative transfer and statistical equilibrium equations
for boron were solved with a revised version of the DETAIL program
\citep{detail} based on the accelerated lambda iteration following the
efficient method described by \citet{rybicki1991,rybicki1992}. In
addition to the continuous background opacities, line opacities
introduced by hydrogen and metal lines calculated with OS method were
also taken into account in solving the equations. Lines shortward and
longward of 2500\,{\AA} were extracted from the VALD database
\citep{vald} and the line list of \citet{kurucz1992}, respectively.
Please refer to \citet{shi2008} for more details about the dealing
with background opacities.

\begin{figure}
\centering
\includegraphics{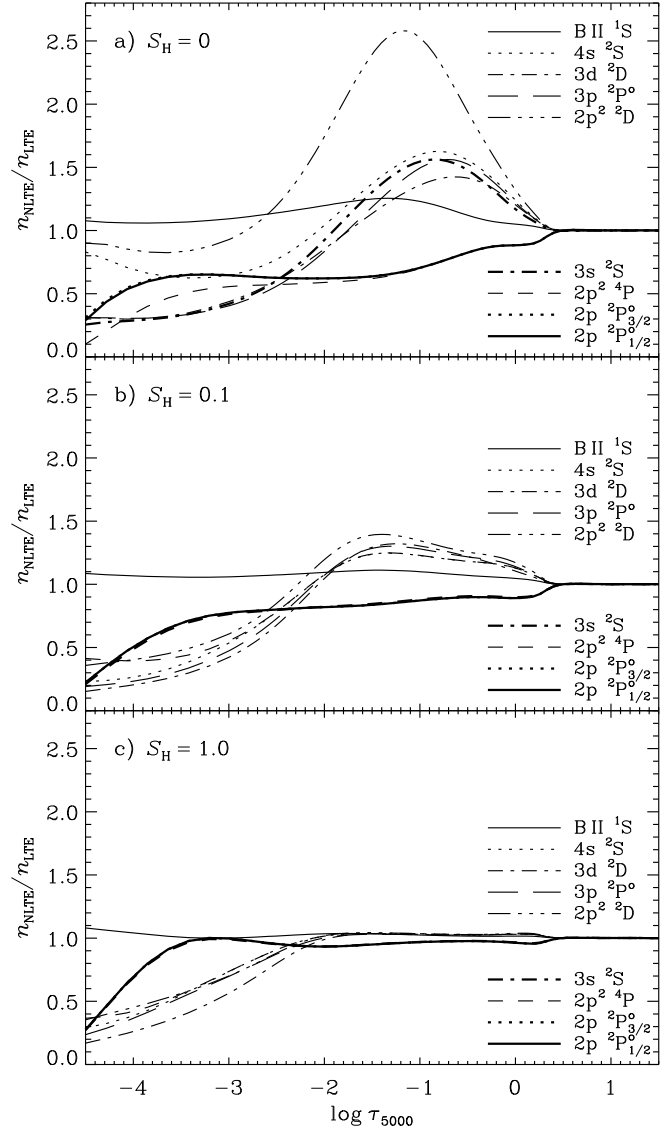}
\caption{Departure coefficients for the populations of first eight
energy levels of \ion{B}{1} and the ground state of \ion{B}{2} for the
representative metal-poor star HD\,140283 calculated with the stellar
parameters from K94.\label{fig:depature}}
\end{figure}

The departures from LTE for the populations of the first eight energy
levels of \ion{B}{1} and the ground state of \ion{B}{2} calculated
with different collision strengths with neutral hydrogen are shown in
Figure~\ref{fig:depature} for the representative metal-poor star
HD\,140283 (note that, for comparison, the departure coefficients
shown here is calculated with the the same stellar parameters as K94,
while the final abundances of HD\,140283 are determined with the
stellar parameters given in Table~\ref{table:abun}). It can be seen in
Figure~\ref{fig:depature}a that our results confirm the conclusions of
K94 that the departures from LTE for the resonance lines are due to a
combination of {\it overionization} and {\it optical pumping} effects.
However, the amplitude of the departures from LTE for our results (see
Figure~\ref{fig:depature}a) are smaller than that of K94 (c.f.
Figure~4 of K94). As described in $\S$~\ref{atom}, the main
differences in the boron atomic model between K94 and this work lie in:
\begin{itemize}
\item excitation energies for some of the higher levels;
\item number of radiative $b$-$b$ transitions considered;
\item photoionization cross-sections for the $^4\mathrm{P}$ and
$^2\mathrm{F^o}$ levels;
\item electron impact excitation cross-sections;
\item cross-sections of ionization induced by collisions with neutral
hydrogen.
\end{itemize}
The marginal differences in excitation energies and photoionization
cross-sections of the higher levels cannot produce such large
differences in the NLTE effects. The effects of the number of
radiative $b$-$b$ transitions considered was tested by excluding the
transitions with $f<0.001$ as K94 did, but calculated results showed
that there were no obvious differences. The differences in NLTE
effects caused by the differences in electron impact excitation
cross-sections was investigated in an extreme way, i.e., calculation
was performed without the electron impact processes; in this case, the
departures from LTE for our results increase a bit, but still are
smaller than that of K94. Lastly, we checked the influences of the
hydrogen collisional ionization on the NLTE effects of \ion{B}{1} with
an elaborate routine, which enabled us to treat the hydrogen
collisional excitation and ionization separately. We first fixed the
strength of hydrogen collisional excitation, and then calculated the
NLTE effects with two different strengths of hydrogen collisional
ionization ($S_{\mathrm{H}}=0$ and 1, respectively). As a result, only
marginal differences in the NLTE effects were found, which means that
hydrogen collisional ionization process plays an insignificant role in
the \ion{B}{1} line formation. Similar conclusion was also derived by
K94, who found the hydrogen collisional ionization rates to be several
orders of magnitude smaller than the corresponding rates due to
electron collisions. In a word, it's unlikely that the large
differences in the NLTE effects between K94 and this work are due to
the differences in atomic model.

As pointed by K94, both the overionization and optical pumping effects
are caused by the hot, non-local, ultraviolet radiation fields in the
line formation regions, while the background opacities in the NLTE
calculation code used by K94 were just the continuous ones, i.e., the
background line opacities were not included. To compensate this defect,
K94 approximated the line blanketing effects on the photoionization
processes by treating the transitions at fixed rates which were
calculated from a specified radiation field. These radiation fields
were produced directly by the adopted model atmosphere code, which had
included the line blanketing via opacity sampling. However, such
approximation was not applied to the radiative $b$-$b$ transitions
(except for some test runs), therefore the efficiency of the pumping
processes would be exaggerated. In their subsequent work, KC96 used a
revised version of NLTE calculation code which allows the inclusion of
spectral lines in the background opacities. As a result, the
departures from LTE decreased. For example, the NLTE abundance
correction ($\Delta\log\epsilon\mathrm{(B)}\footnote{$\log\epsilon
\mathrm{(X)}=\log[N\mathrm{(X)}/N\mathrm{(H)}]+12$}=\log\epsilon
\mathrm{(B)_{NLTE}}-\log\epsilon\mathrm{(B)_{LTE}}$) for the \ion{B}{1}
2496.771\,{\AA} resonance line decreased from 0.59 to 0.49\,dex for
HD\,140283 after including background line opacities. However, the
departures from LTE for the KC96's results are still larger than ours
(as will be presented below, our NLTE B abundance correction for
HD\,140283 amounts only to 0.2\,dex when neglecting collisions with
neutral hydrogen). The NLTE calculation code used in this work has
included most of the newest continuous and line opacities, while some
of them (the \ion{Fe}{1} $b$-$f$ opacities of \citealt{bautista1997}
and the VALD atomic line data being the most prominent) were not
available for KC96 more than one decade ago. It might be possible that
KC96 used smaller background opacities than that adopted in this work
and thus derived larger NLTE effects for \ion{B}{1}.

Another difference between the results of K94 and this work worthwhile
to be noted is that K94 concluded that the departures from LTE would
{\it increase} if including the $b$-$b$ inelastic collisions with
neutral hydrogen. This is completely contrary to our results that the
line formation of \ion{B}{1} will get closer to LTE if the strength of
collisions with neutral hydrogen increases, as can be seen clearly in
Figure~\ref{fig:depature}. As the role of the hydrogen collisional
ionization has been shown to be insignificant, the obvious changes in
the NLTE effects should be mostly due to the hydrogen collisional
excitation.

\begin{figure}
\centering
\includegraphics{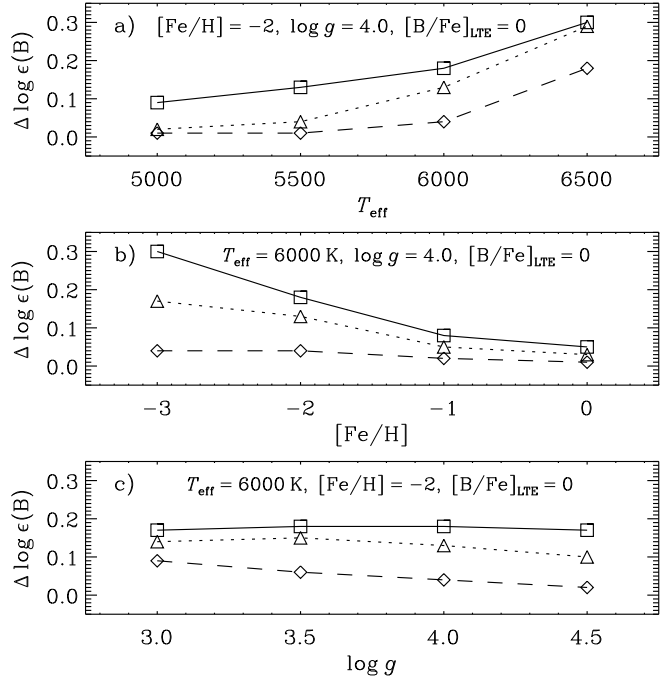}
\caption{Dependence of NLTE abundance corrections for the \ion{B}{1}
2496.771\,{\AA} resonance line on (a) effective temperature, (b)
metallicity, and (c) surface gravity. Squares, triangles, and diamonds
represent results calculated with different strength of collisions
with neutral hydrogen, i.e., $S_{\mathrm{H}}=0$, 0.1, and 1
respectively.\label{fig:nlte}}
\end{figure}

Figure~\ref{fig:nlte} shows the NLTE abundance corrections for the
\ion{B}{1} 2496.771\,{\AA} resonance line as a function of stellar
parameters. In general, the NLTE effects are large for hot,
metal-poor, and low-gravity stars. This is in agreement with the
results of KC96.

\section{Abundances and uncertainties}\label{results}

\subsection{Boron}

\begin{figure}
\centering
\includegraphics{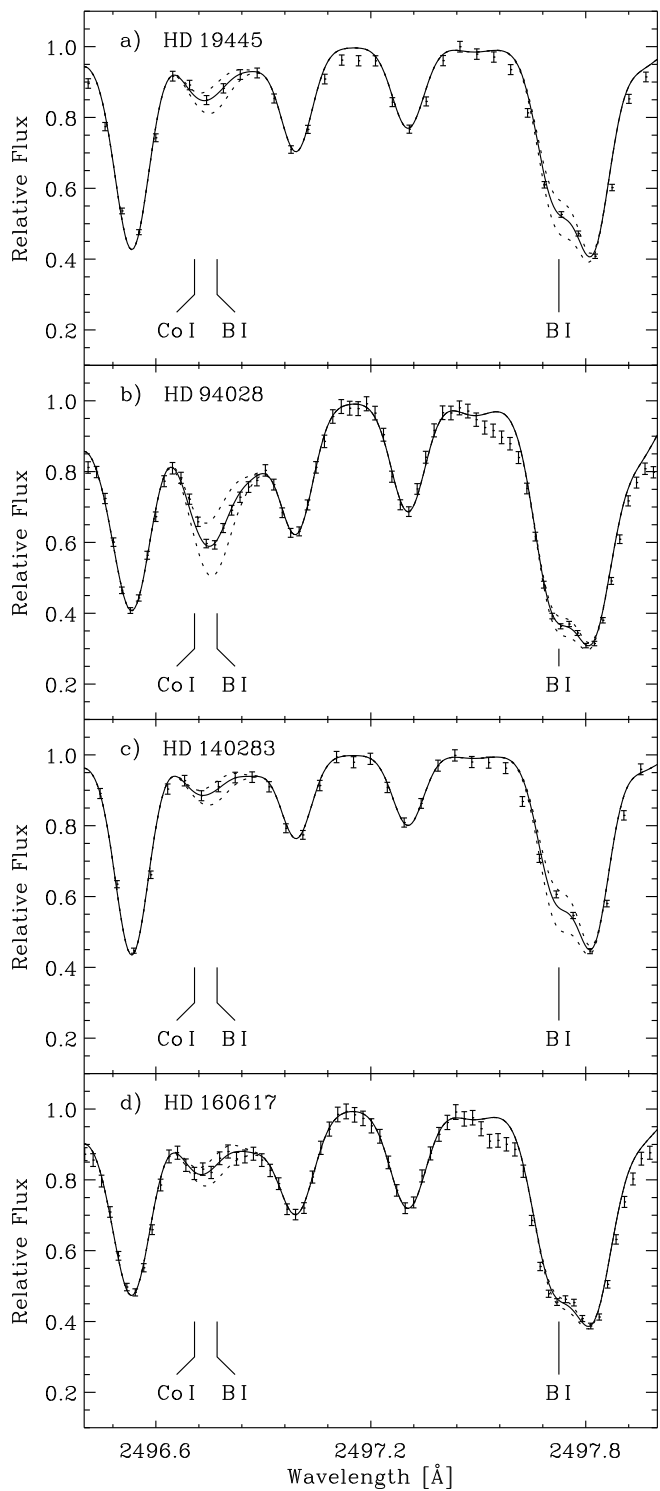}
\caption{Spectrum synthesis of the \ion{B}{1} 2497\,{\AA} resonance
doublet for four program stars. The observed spectra are represented
by photon statistics error bars. The solid line is the best-fitting
synthesis, and the dotted lines are synthetic spectra with B
abundances of $\pm0.2$\,dex relative to the best synthesis.\label{fig:synth}}
\end{figure}

B abundances were derived by spectrum synthesis of the \ion{B}{1}
2497\,{\AA} resonance doublet using the IDL/FORTRAN SIU software
package developed by Dr.~J.~Reetz. The input line list was extracted
from \citet{duncan1998a} with minor changes. We took the weaker
2496.771\,{\AA} component as the primary abundance indicator because
the stronger 2497.725\,{\AA} line is severely blended. Though the
weaker line is also blended with a \ion{Co}{1} line at 2496.708\,{\AA},
it does not have significant effect on the derived B abundance as
shown by \citet{duncan1997} and \citet{primas1999}. Finally, we
determined B abundances for 12 sample stars; for the rest four stars,
as reported by \citet{garcialopez1998}, a combination of high
effective temperature ($T_{\mathrm{eff}}>6000$\,K) and low metallicity
($\mathrm{[Fe/H]}<-2$) prevented us from obtaining accurate B
abundances with the \ion{B}{1} 2496.771\,{\AA} line, thus only upper
limits were given. Figure~\ref{fig:synth} shows the observed and
synthetic spectra of the \ion{B}{1} 2497\,{\AA} resonance doublet for
four program stars.

The uncertainties associated with B abundances were estimated
following the same procedure described by \citet{duncan1997}.
Individual errors caused by uncertainties in effective temperature,
surface gravity, metallicity, microturbulent velocity, photon
statistics in observed spectrum, and continuum placement (a typical
value of $\pm2$\% was adopted) were added in quadrature. The B
abundances and final errors are given in Table~\ref{table:abun}.

\begin{figure}
\centering
\includegraphics{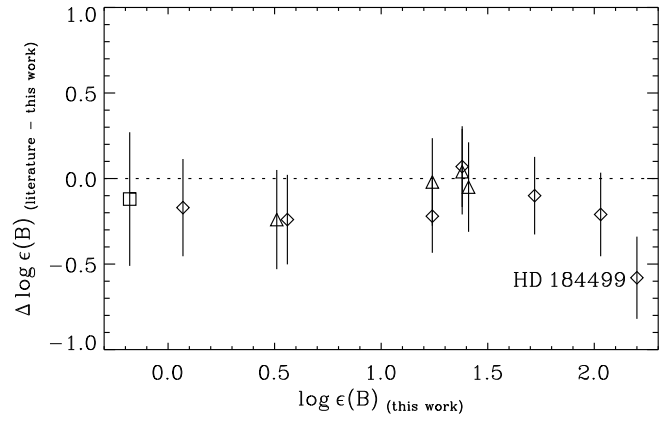}
\caption{Comparison of B abundances (LTE) between the literatures and
this work for the 12 stars with determined B abundances: diamonds,
\citet{duncan1997}; square, \citet{garcialopez1998}; triangles,
\citet{primas1999}.\label{fig:b_comp}}
\end{figure}

As mentioned above, B abundances of our sample stars have been studied
previously by other investigators \citep{duncan1997,garcialopez1998,
primas1999}. Figure~\ref{fig:b_comp} shows the comparison of LTE B
abundances between the literatures and this work for the 12 stars with
determined B abundances. It can be seen that the agreement is
reasonable for most of the stars. The only exception is HD\,184499,
for which our B abundance is 0.58\,dex higher than that
of \citet{duncan1997}. Such a large difference is mostly due to the
different stellar parameters adopted by \citet{duncan1997} and by this
work (their effective temperature and metallicity are 120\,K and
0.4\,dex lower than ours, respectively). If we adopted the same
stellar parameters as \citet{duncan1997}, then the difference in B
abundance will decrease from 0.58 to 0.28\,dex, which is comparable to
the uncertainties.

\subsection{Oxygen}

As it is believed that boron production is related to oxygen directly,
O abundance would be a preferred alternative to Fe in investigating
the origin and evolution of B in the Galaxy. There are several
indicators for O abundance: the UV OH lines, the [\ion{O}{1}] 6300
and 6363\,{\AA} forbidden lines, the \ion{O}{1} 7774\,{\AA} triplet,
and the infrared (IR) vibration-rotation OH lines, and different
indicators may give different O abundances. As discussed in
\citet{nissen2002} and references therein, O abundances from the OH
and \ion{O}{1} lines are very sensitive to the adopted stellar
parameters, such as the effective temperature and surface gravity;
moreover, line formations in cool metal-poor stars are probably far
from LTE for either the UV OH lines or the \ion{O}{1} triplet. In
contrast, [\ion{O}{1}] forbidden lines are formed very close to LTE
and O abundances from these lines are not sensitive to the adopted
stellar parameters. Therefore, it is normally believed that
[\ion{O}{1}] forbidden lines are the most reliable indicators for
O abundances, but the difficulty is that these lines are very weak in
metal-poor dwarf and subgiant stars.

For our sample stars, in the absence of reliable data for the
[\ion{O}{1}] forbidden lines, we used the \ion{O}{1} triplet to
determine their O abundances. Based on the equivalent widths of the
\ion{O}{1} 7774\,{\AA} triplet collected from the literatures, O
abundances were first calculated with $\log gf=0.369$, 0.223, and
0.002 from \citet{wiese1996} in LTE. Then NLTE corrections were
applied according to the results of \citet{fabbian2009} which were
calculated with new electron collisional data from \citet{barklem2007}.
We were aware that the results of \citet{fabbian2009} were based on
the MARCS models \citep{gustafsson2008}, while our LTE O abundances
were determined with the MAFAGS-OS9 models. However, we also noted
that \citet{fabbian2009} found that, when using the ATLAS9 models of
\citet{castelli2003}, resulting NLTE abundance corrections are very
similar to that obtained using MARCS models for intermediately low
metallicity ($\mathrm{[Fe/H]}>-2.5$) stars. This is not unexpected
because of the the similarity between ATLAS9 and MARCS models. As the
MAFAGS-OS9 model used in this work is also quite similar to the MARCS
model atmospheres\footnote{We have made comparisons between MAFAGS-OS9
and MARCS for a set of model atmospheres across the parameter space
investigated here, and results show that the temperature, electron
density, and gas pressure stratifications of the MAFAGS-OS9 models are
very similar to that of the MARCS models.}, we suggest that the NLTE
corrections derived by \citet{fabbian2009} can be applied to our
results too. The NLTE corrections of \citet{fabbian2009} were
calculated with two different hydrogen collisional strengths, namely
$S_{\mathrm{H}}=0$ and 1. The latter was adopted in this work because
both \citet{allendeprieto2004} and \citet{pereira2009} found that the
$S_{\mathrm{H}}=1$ case could reproduce the solar center-to-limb
variations better than $S_{\mathrm{H}}=0$. Moreover, as stated by
\citet{fabbian2009}, in the case of $S_{\mathrm{H}}=1$, derived [O/Fe]
trend becomes almost flat below $\mathrm{[Fe/H]}\sim-1$, which is in
better agreement with the results obtained form other O abundance
indicators; and if $S_{\mathrm{H}}=0$ is adopted, then [O/Fe] will
decrease towards lower [Fe/H], which would open new questions. The
final NLTE-corrected O abundances are given in Table~\ref{table:abun}
(the reference solar O abundance is $\log\epsilon (\mathrm{O})=8.77$
from \citealt{tan2009}; we emphasize that the slopes of the Be(B)-O
abundance relationships, which are of our primary concern, are
{\it independent} of the choice of the reference solar O abundance).

The uncertainties of O abundances were estimated from the errors of the
stellar parameters and the equivalent widths. Typical errors of
$\pm$80\,K in effective temperature and $\pm$0.15\,dex in surface
gravity both lead to an error of $\pm$0.05\,dex in O abundance. The
uncertainties in iron abundance and microturbulent velocity nearly
have no effect on O abundance. A typical error of $\pm$3\,m{\AA} in
equivalent width corresponds to an error of $\pm$0.04\,dex. In total,
the typical error of O abundance is around $\pm$0.08\,dex.

\subsection{Beryllium}

Beryllium abundance can be a useful tool to understand the origin and
evolution of boron. It is suggested that Be is only produced by
spallation reactions, while B may be additionally yielded by the
$\nu$-process, therefore, the B/Be ratio can put some constraints on
the production history of B. We computed the Be abundances from the
equivalent widths of the \ion{Be}{2} 3131.066\,{\AA} resonance line
($\log gf=-0.468$ from \citealt{kurucz1992} was adopted) given by
\citet{boesgaard1993} and \citet{boesgaard1999} for 13 of our sample
stars. For the remaining three stars, Be abundances were adopted from
the literatures but scaled to our stellar parameters. Beryllium
abundances of all the stars were determined in LTE as NLTE effects for
the \ion{Be}{2} resonance lines are insignificant according to the
results of \citet{garcialopez1995}.

The uncertainties of Be abundances mainly come from the errors of the
stellar parameters and equivalent widths. The uncertainties associated
with the equivalent widths, which were not given by
\citet{boesgaard1993} and \citet{boesgaard1999}, depend mostly on the
placement of continuum. As discussed by \citet{duncan1997} for boron,
the continuum is easier to define for more metal-poor stars, but the B
lines are weaker; for stars with higher metallicity, the errors in
continuum placement are larger but the B lines are stronger, so the
combined effects are relatively constant over the metallicity range
investigated. This situation holds also for Be. We noted that
\citet{boesgaard1993} claimed a typical relative error of about 1\% (in
the worst cases 2\%) for their continuum placement, and \citet{boesgaard1999}
estimated the typical Be abundance errors due to the placement of the
continuum to be 0.02--0.06\,dex. In this work, we adopted a conservative
value of 0.1\,dex as the typical uncertainty caused by the errors in
continuum location. The Be abundances and final errors are listed in
Table~\ref{table:abun}.

\section{Discussion}\label{discuss}

\begin{figure}
\centering
\includegraphics{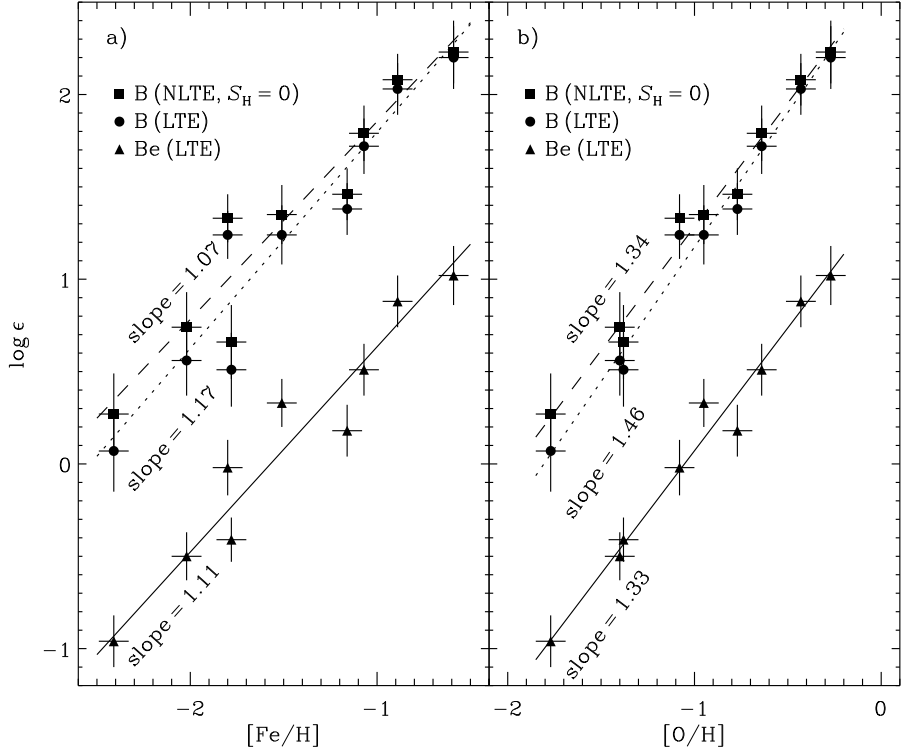}
\caption{Be and B abundances against (a) [Fe/H] and (b) [O/H] for the
nine Be- and B-normal stars. Filled triangles are Be abundances,
filled circles are LTE B abundances, and filled squares are NLTE B
abundances calculated without neutral hydrogen collisions
($S_{\mathrm{H}}=0$). Solid, dotted, and dashed lines are the best
linear fits to the Be-Fe(O), $\mathrm{B_{LTE}}$-Fe(O), and
$\mathrm{B_{NLTE,S_{H}=0}}$-Fe(O) relationships respectively.\label{fig:fe_o_be_b}}
\end{figure}

In this section we will investigate the implications for the origin
and evolution of boron from the abundances of our sample stars. Before
that it is necessary to check whether some of the stars are depleted
in B because B can be destroyed in stars by fusion reactions at a
relatively low temperature (about $5\times10^6$\,K). This can be done
by investigating their Li and Be abundances because the destruction
temperature for Li, Be, and B increases in order. If Li or Be is not
depleted, then B should not be depleted either. Among the 12 stars
with determined B abundances, five are very likely depleted in Li (far
below the Spite plateau) according to \citet{garcialopez1998} and
\citet{primas1999}. In the following some comments are given for these
Li-depleted stars:
\begin{itemize}
  \item {\it HD\,64090 \& HD\,184499}: though these two stars are
  obviously depleted in Li, their Be and B abundances match well with
  the Be-Fe(O) and B-Fe(O) trends (abundance trends in logarithmic
  plane, this meaning holds true for all the following denotations in
  the same style if not specified) determined by the Li-normal (thus
  Be- and B-normal) stars as can be seen in Figure~\ref{fig:fe_o_be_b}.
  We conclude that their Be and B abundances are not depleted and will
  take them into account when investigating the origin and evolution
  of B.
  \item {\it HD\,106516 \& HD\,221377}: both of them have only upper
  limits for their Li and Be abundances. Their B abundances are
  slightly lower than that of stars with similar metallicity.
  \citet{primas1999} also took these two stars as possible B-depleted
  stars. We will exclude them in the following discussion.
  \item {\it BD\,+23$\degr$3130}: this is the most metal-poor star
  with determined B abundance in our sample. It is possibly a giant
  star with $T_{\mathrm{eff}}=5255$\,K and $\log g=2.89$. Its Be and B
  abundances are 0.37 and 0.25\,dex lower, respectively, than that of
  HD\,140283, which has very similar metallicity. It's hard to say
  whether this star is slightly depleted in Be or B, but for
  reliability, it will not be considered either in the following
  investigation.
\end{itemize}
Another star worthy of being noted is HD\,160617, which was concluded
as a Li-normal but B-depleted star (whether it is Be-depleted depends
on the adopted stellar parameters) by \citet{primas1998}. Both the Li
and Be abundances of this star determined with our stellar parameters
seem to be normal \citep[see][]{tan2009}. Its B abundance obtained by
this work is 0.24\,dex higher than that of \citet{primas1998} which
was determined with very similar parameters, but considering the
uncertainties, they are still in reasonable agreement. However, we do
note that Be and B abundances of HD\,160617 are much smaller than that
of HD\,64090 which has very similar Fe abundance. This large
differences lead to the relatively large scatter at
$\mathrm{[Fe/H]}\sim-1.8$ on the Be-Fe and B-Fe trends as can be seen
in Figure~\ref{fig:fe_o_be_b}a. But on the Be-O and B-O diagram, the
scatter is much smaller as shown in Figure~\ref{fig:fe_o_be_b}b. This
can be a direct evidence that Be and B production are directly related
to O rather than Fe. Therefore, we take HD\,160617 as a B-normal star
and will include it in the following investigation on the B production
and evolution. Finally, we have nine stars with normal Be and B
abundances and the following discussion will be only based on these
stars.

\subsection{Evolution of Be and B with metal abundances}

Figure~\ref{fig:fe_o_be_b} shows the Be and B abundances against
[Fe/H] and [O/H] for the nine Be- and B-normal stars. It is obvious
that both $\log\epsilon\mathrm{(Be)}$ and $\log\epsilon\mathrm{(B)}$
increase approximately linearly with increasing [Fe/H] or [O/H].
Slopes and intercepts for the linear leat-squares fits to the
abundance data are given in Table~\ref{table:fit}; for comparison,
results from previous studies are also given. It can be seen that the
slope of the Be-Fe trend obtained by this work agrees well with the
result of \citet{tan2009}, and is consistent with that of
\citet{smiljanic2009} within error bars. But our B-Fe trend is
marginally steeper than that of \citet{boesgaard1999} and
\citet{rich2009}, which is mainly due to the different metallicity
ranges investigated. When the fitting is only restricted to the
similar metallicity range to this work, the slopes of the Be-Fe trends
from \citet{boesgaard1999} and \citet{rich2009} are 1.17 and 1.07,
respectively, which are consistent with our result. For the B-Fe
trend, the slope obtained by this work is steeper than that of
\citet{duncan1997}. This difference can be attributed to the two
lowest metallicity stars (BD\,$+3\degr 740$ and BD\,$-13\degr 3442$)
in the sample of \citet{duncan1997}, for which only upper limits of B
abundances are derived by this work, and thus are not included in our
investigation of the B-Fe(O) relationships. It can be seen clearly in
Figure~3 of \citet{duncan1997} that, compared with the general B-Fe
trend, these two stars show obvious enhancement of B abundances, and
if they are excluded from the fitting, then the slope of the B-Fe
trend will increase to 1.04, which is consistent with our result
within uncertainties. For the Be-O and B-O trends, comparisons become
more complicated because different work may use different O abundance
indicators and/or different descriptions of the line formation (LTE
versus NLTE). The slope of the Be-O trend obtained by this work agrees
well with the result of \citet{tan2009}, of which the O abundances
were also determined using the \ion{O}{1} triplet and had been
corrected for NLTE effects. \citet{boesgaard1999} adopted the average
values of O abundances based on the UV OH lines and the \ion{O}{1}
triplet; \citet{rich2009} determined O abundances from the UV OH
lines, therefore, a direct comparison of their results with that of
this work is inappropriate. \citet{smiljanic2009} used [$\alpha$/H] as
a proxy for [O/H] in the absence of O abundances for all their sample
stars, but interestingly, the Be-$\alpha$ trend obtained by them
agrees well with the Be-O trend of this work. Oxygen abundances
adopted by \citet{duncan1997} for most of their sample stars were
based on the \ion{O}{1} triplet and had been corrected for NLTE
effects, but for the same reason as we have mentioned in the
comparison of the B-Fe trend, their B-O trend is flatter than that of
this work.

Both the slopes of our Be-Fe and Be-O trends are closer to unity,
which confirms the previous conclusions that Be should be mainly
produced by primary process in the early Galaxy. As for B, the derived
B-Fe and B-O relationships depend somehow on the adopted strength of
collisions with neutral hydrogen as we have discussed in
$\S$~\ref{nlte_result}. As shown in Figure~\ref{fig:nlte}b, on one
side, NLTE abundance corrections will increase if the strength of
collisions with neutral hydrogen decreases; on the other side, NLTE
abundance corrections increases with decreasing metallicity, therefore,
the $\mathrm{B_{LTE}}$-Fe(O) and $\mathrm{B_{NLTE,S_{H}=0}}$-Fe(O)
trends shown in Figure~\ref{fig:fe_o_be_b} represent the two extreme
cases, and the real situation should lie between them. In other words,
the slopes of the B-Fe and B-O relationships should be restricted to
the intervals [1.07, 1.17] and [1.34, 1.46], respectively. In any
case, the slopes of the B-Fe and B-O trends are closer to 1, which
indicates that primary process is dominant in B production in the
early Galaxy. In addition, there seems to be slight changes in slopes
for both the Be-Fe(O) and B-Fe(O) trends around $\mathrm{[Fe/H]}\sim-1$
and $\mathrm{[O/H]}\sim-1$, respectively, which was also reported by
\citet{garcialopez1998}, but there are too few stars in our sample to
confirm it.

It is believed that both Be and B can be produced by spallation
reactions while B may be additionally produced by the $\nu$-process,
therefore, B should increase more rapidly than Be (or at a similar
speed as Be if the contribution of the $\nu$-process is neglectable).
However, as first pointed by \citet{boesgaard2004}, after applying the
NLTE abundance corrections of KC96, the B abundance increases more
slowly for halo stars than Be does (see Tables~5 and 6 of
\citealt{duncan1997} and Figures~11, 13, and 14 of
\citealt{boesgaard2004}), which is in contradiction with the
theoretical models. For our results, as shown in
Figure~\ref{fig:fe_o_be_b}, this problem {\it no longer} exists no
matter the strength of collisions with neutral hydrogen may be. Even
in the extreme case (no collisions with neutral hydrogen), B increases
with Fe or O at a very similar speed as Be does, and if the strength
of collisions with neutral hydrogen increases, then B will increase
with Fe or O a bit faster than Be. To further investigate whether our
new result (B increase with Fe or O in a similar fashion to Be) are
due to our smaller NLTE abundance corrections or to the LTE B
abundances, we applied the NLTE abundance corrections of KC96 to our
LTE B abundances. As a result, the slopes of the B-Fe and B-O trends
decrease to 0.97 and 1.22, respectively, which are lower than the
slopes of the Be-Fe and Be-O trends. In addition, a reverse test was
also performed, i.e., we applied our NLTE corrections to the LTE B
abundances of \citet{duncan1997}, and in this case, the slope of the
B-Fe trend would be 0.86 (0.70 when the NLTE corrections of KC96 are
applied), which is still similar to the slope of the Be-Fe trend
(0.85). Therefore, our result that B increases with Fe or O at a very
similar speed as Be is mostly due to the smaller NLTE abundance
corrections.

Though good linear relationships between Be and Fe as well as between
B and Fe are found for the metallicity range ($-2.5<\mathrm{[Fe/H]}<-0.5$)
investigated here, it is not clear whether such trends will continue
to lower metallicities. \citet{primas2000} reported the detection of
Be in the very metal-poor star G\,64--12 ($\mathrm{[Fe/H]}=-3.3$) and
found that Be abundance of this star is significantly higher than what
expected from the general Be-Fe trend. This led them to suggest a
possible flattening of Be abundances at very low metallicities. We
also note that, for BD\,$+3\degr740$ and BD\,$-13\degr3442$, only
upper limits of B abundances are found by \citet{garcialopez1998} and
by this work, while \citet{duncan1997,duncan1998b} claimed the
detection of B in these two stars. And interestingly, B abundances of
BD\,$+3\degr740$ and BD\,$-13\degr3442$ given by \citet{duncan1997,
duncan1998b} also seem to be higher than what expected from the
general B-Fe trend (both of these two stars are at the metallicity end
of their sample). Moreover, \citet{asplund2006} reported the detection of \element{Li}{6}
(at the significance level $\geq2\sigma$) in nine metal-poor stars.
Their most noteworthy result is the detection of an abnormally high
\element{Li}{6} abundance in the very metal-poor star LP\,815--43,
which cannot be explained either by the standard BBN or by the
Galactic cosmic-ray spallation and $\alpha$-fusion reactions. One
common solution to all the three problems described above can be
non-standard BBN, which could produce much more primordial light
elements than the standard BBN. However, if ignoring the problem of
high \element{Li}{6} abundance at very low metallicity (whether
\element{Li}{6} is really detected in some metal-poor stars is still
disputable, for example, \citealt{garciaperez2009} aimed to determine
\element{Li}{6}/\element{Li}{7} for a sample of metal-poor stars, and
only found that the \element{Li}{6}/\element{Li}{7} ratios are
comparable to or even lower than the associated uncertainties), there
may be no need to invoke the non-standard models of BBN to explain the
possible Be and B ``plateau'' as pointed by \citet{vangioni1998}. They
proposed that in the very early Galaxy, if supernovae of all masses
paly roles in the production of Be(B), then Be(B) should continue to
decrease with decreasing [Fe/H]; otherwise, if only the most massive
stars (initial mass larger than 60\,$M_{\sun}$) contribute to the
Be(B) yields, then Be(B) should be roughly constant. They suggested
that the predicted difference in the behavior of Be(B) could be tested
by observations at $\mathrm{[Fe/H]}\leq-3$. Recently, \citet{rich2009}
investigated the Be abundances of 24 very metal-poor stars with [Fe/H]
from $-2.3$ to $-3.5$ and found changes in slopes for the Be-Fe trend
at $\mathrm{[Fe/H]}\sim-2.2$ as well as for the Be-O trend at
$\mathrm{[O/H]}\sim-1.6$. In general, their results show that Be
increases with Fe(O) much more slowly below $\mathrm{[Fe/H]}\sim-2.2$
($\mathrm{[O/H]}\sim-1.6$) than above that. Though there is no
indication of a Be plateau, the four most metal-deficient stars
($\mathrm{[Fe/H]}<-3$) do show a Be enhancement with respect to Fe at
the $1\sigma$ level, which is qualitatively consistent with the
predictions of the light-element production model involved with only
the most massive stars proposed by \citet{vangioni1998}. To clarify
this problem, more accurate data on Be and B abundances at very low
metallicities are needed.

\subsection{The B/Be ratio}

\begin{figure}
\centering
\includegraphics{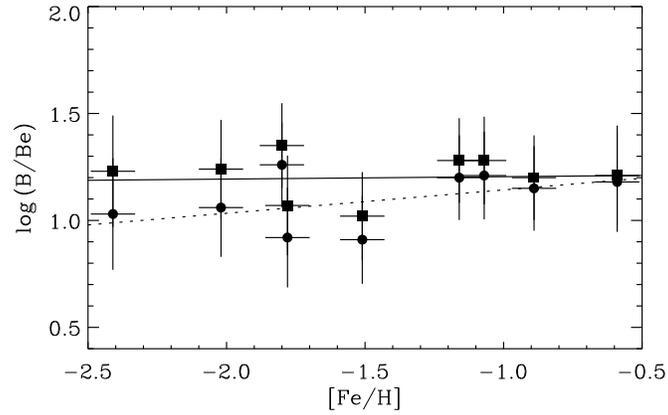}
\caption{B-to-Be abundance ratio as a function of [Fe/H] for the nine
Be- and B-normal stars (filled circles: B abundances calculated in LTE;
filled squares: B abundances calculated in NLTE without neutral
hydrogen collisions, i.e., $S_{\mathrm{H}}=0$). Dotted and solid lines
represent the best linear fits to the $\mathrm{B_{LTE}}$/Be-Fe and
$\mathrm{B_{NLTE,S_{H}=0}}$/Be-Fe relationships, respectively.\label{fig:be_b}}
\end{figure}

Figure~\ref{fig:be_b} shows the B/Be ratio as a function of
metallicity for the nine Be and B-normal stars. Linear least-squares
fits to the data give the following relationships:
\begin{eqnarray*}
\log\mathrm{(B_{LTE}/Be)}=(0.10\pm0.14)\mathrm{[Fe/H]}+(1.25\pm0.21)\\
\log\mathrm{(B_{NLTE,S_{H}=0}/Be)}=(0.01\pm0.12)\mathrm{[Fe/H]}+(1.22\pm0.19)
\end{eqnarray*}
It is clear that, in LTE case, B/Be increases slightly with increasing
metallicity, while in the case of NLTE but without neutral hydrogen
collisions, B/Be is almost constant. Again, the real situation for the
B/Be-Fe trend should lie between the two extreme cases, but
considering the uncertainties, the B/Be ratio is roughly constant over
the metallicity range investigated here. The average B/Be ratio for
the nine stars should fall in the interval $[13\pm4, 17\pm4]$, of
which the lower and upper limits correspond to the two extreme cases
respectively. This is in reasonable agreement with the results
$15\pm3$ of \citet{duncan1997} and $20\pm10$ of \citet{garcialopez1998}.
Our average B/Be ratio is consistent with the predicted value of
spallation process, which depends on the compositions and energy
spectra of cosmic rays (for example, a typical spallation-produced B/Be
ratio of about 14 is derived by \citealt{ramaty2000}). However, if
assuming that B is purely produced by spallation reactions and the
\element{B}{11}/\element{B}{10} ratios in metal-poor stars are the
same as the meteoric value, then a minimum B/Be ratio of about 20 is
required \citep{vangioni1996}. This value is slightly larger than our
result, which might indicate that B is also produced by the
$\nu$-process. \citet{ramaty2000} showed that even though the
production histories of the spallation-produced B and Be and the
neutrino-produced B are different, B/Be can still remain essentially
constant as a function of [Fe/H]. However, the \element{B}{11}/\element{B}{10}
ratios in metal-poor stars need not to be the same as the meteoric
value, especially considering the fact that our average B/Be ratio in
metal-poor stars is lower than the solar meteoric value of about 23.
Anyway, an accurate measurement of the \element{B}{11}/\element{B}{10}
ratios in metal-poor stars is strongly desirable.

\section{Summary}\label{summary}

In this work, we have investigated the NLTE line formation of neutral
boron in the atmospheres of cool stars. Our results confirm the
conclusions of K94 and KC96 that NLTE effects for the \ion{B}{1}
resonance lines, which are due to a combination of overionization and
optical pumping effects, are significant for hot, metal-poor, and
low-gravity stars. However, the amplitude of the departures from LTE
found by the present study are smaller than that of KC96, which might
be because that KC96 used smaller background opacities than that
adopted in this work. In addition, our calculation shows that
departures from LTE for the \ion{B}{1} resonance lines decrease with
increasing strength of collisions with neutral hydrogen, which is
completely contrary to the result of K94.

The NLTE line formation results for \ion{B}{1} have been applied to
the determination of B abundances for a sample of 16 metal-poor stars,
for which the Be and O abundances are also determined. No matter the
strength of collisions with neutral hydrogen may be, both the slopes
of the Be-O and B-O trends determined by the nine Be- and B-normal
stars are closer to unity, which confirms the previous conclusions
that primary process is dominant in the production of Be and B in the
early Galaxy. Most importantly, our results show that, B increases
with Fe or O at a very similar speed as, or a bit faster than Be does,
which is in accord with the theoretical models. The B/Be abundance
ratios are roughly constant over the metallicity range
$-2.5<\mathrm{[Fe/H]<-0.5}$. The average B/Be ratio falls in the
interval $[13\pm4, 17\pm4]$, which is consistent with the predictions
of spallation process. However, if assuming the isotopic
\element{B}{11}/\element{B}{10} ratios in metal-poor stars to be the
same as the meteoric value, then a minimum B/Be ratio of about 20 is
required by the pure spallation process, therefore the $\nu$-process
may be needed to reconcile this problem. But whether the
\element{B}{11}/\element{B}{10} ratios in metal-poor stars are the
same as the meteoric value is still unclear, thus an accurate
measurement is strongly desirable, though rather challenging.

\acknowledgments

We are grateful to Dr.~Frank Grupp for providing the MAFAGS-OS9 model
atmosphere code. K.T. would like to thank Dr.~Martin Asplund for
comments on the manuscript and suggestions on the NLTE abundance
corrections for oxygen. Thanks also go to the anonymous referee for
the valuable suggestions and comments. This work is supported by the
National Nature Science Foundation of China under grant Nos.~10778626,
10821061, 10973016 and by the National Basic Research Program of China
under grant No.~2007CB815103. This research has made use of the SIMBAD
database, operated at CDS, Strasbourg, France.

\clearpage

\clearpage

\clearpage

\begin{deluxetable}{lccccr}
\tablewidth{0pt}
\tablecolumns{6}
\tablecaption{Stellar parameters\label{table:par}}
\tablehead{\colhead{Star} & \colhead{$T_{\mathrm{eff}}$} & \colhead{$\log g$} & \colhead{[Fe/H]} & \colhead{$\xi$} & \colhead{Ref.}\\
                          & \colhead{(K)} & \colhead{(cgs)} & \colhead{(dex)} & \colhead{(km\,s$^{-1}$)} & }
\startdata
HD\,19445	          & 6009 & 4.40 & $-2.02$ & 1.6 & 1, 2\tablenotemark{a}\\
HD\,64090	          & 5525 & 4.57 & $-1.80$ & 2.0 & 5   \\
HD\,76932	          & 5890 & 4.12 & $-0.89$ & 1.2 & 4   \\
HD\,84937	          & 6356 & 4.00 & $-2.16$ & 1.7 & 1, 3\tablenotemark{a}\\
HD\,94028	          & 5925 & 4.19 & $-1.51$ & 1.5 & 2   \\
HD\,106516 	          & 6115 & 4.28 & $-0.80$ & 1.7 & 5   \\
HD\,140283 	          & 5725 & 3.68 & $-2.41$ & 1.5 & 4   \\
HD\,160617            & 5940 & 3.80 & $-1.78$ & 1.5 & 4   \\
HD\,184499 	          & 5820 & 4.07 & $-0.59$ & 1.2 & 5   \\
HD\,194598 	          & 5980 & 4.27 & $-1.16$ & 1.6 & 2   \\
HD\,201891 	          & 5900 & 4.22 & $-1.07$ & 1.2 & 2   \\
HD\,221377 	          & 6330 & 3.81 & $-0.78$ & 2.0 & 5   \\
BD\,$+3^{\circ}740$   & 6290 & 4.02 & $-2.55$ & 1.5 & 2, 3\tablenotemark{a}\\
BD\,$+23^{\circ}3130$ & 5255 & 2.89 & $-2.42$ & 1.7 & 6   \\
BD\,$+26^{\circ}3578$ & 6280 & 3.93 & $-2.25$ & 2.0 & 2   \\
BD\,$-13^{\circ}3442$ & 6390 & 3.88 & $-2.66$ & 1.4 & 3   \\
\enddata
\tablerefs{(1) \citet{korn2003}; (2) \citet{gehren2006}; (3) \citet{mashonkina2008};
(4) \citet{tan2009}; (5) determined with ELODIE spectra by this work; (6) determined
with UVES spectra by this work.}
\tablenotetext{a}{The average value adopted.}
\end{deluxetable}

\begin{deluxetable}{lcrrrcrrrrc}
\tabletypesize{\small}
\tablewidth{0pt}
\tablecolumns{11}
\tablecaption{Abundance results\label{table:abun}}
\tablehead{
\colhead{Star} & \colhead{[O/H]} & \colhead{Ref.} & \colhead{$\log\epsilon$(Be)} & \colhead{Ref.} & \colhead{$\sigma$(Be)}
& \colhead{$\log\epsilon$(B)} & \multicolumn{3}{c}{$\log\epsilon$(B) NLTE} & \colhead{$\sigma$(B)}\\\cline{8-10}
 & \colhead{NLTE} &  & \colhead{LTE} & & & \colhead{LTE} & \colhead{$S_{\mathrm{H}}=0$} & \colhead{$S_{\mathrm{H}}=0.1$} & \colhead{$S_{\mathrm{H}}=1$} & }
\startdata
HD\,19445	         & $-1.40$ & 8  & $-0.50$  & 7 & 0.13    & $ 0.56$ &   0.74  &   0.67  &   0.59  & 0.19   \\
HD\,64090	         & $-1.08$ & 2  & $-0.02$  & 7 & 0.15    & $ 1.24$ &   1.33  &   1.27  &   1.25  & 0.13   \\
HD\,76932	         & $-0.43$ & 11 & $ 0.88$  & 7 & 0.14    & $ 2.03$ &   2.08  &   2.07  &   2.05  & 0.14   \\
HD\,84937	         & $-1.69$ & 2  & $-0.85$  & 7 & 0.12    & $<0.43$ & $<0.73$ & $<0.72$ & $<0.55$ & \nodata\\
HD\,94028	         & $-0.95$ & 4  & $ 0.33$  & 7 & 0.13    & $ 1.24$ &   1.35  &   1.31  &   1.26  & 0.16   \\
HD\,106516 	         & $-0.30$ & 1  & $<-0.58$ & 5 & \nodata & $ 1.38$ &   1.45  &   1.43  &   1.41  & 0.15   \\
HD\,140283 	         & $-1.77$ & 8  & $-0.96$  & 7 & 0.14    & $ 0.07$ &   0.27  &   0.18  &   0.11  & 0.22   \\
HD\,160617           & $-1.38$ & 8  & $-0.41$  & 12& 0.12    & $ 0.51$ &   0.66  &   0.62  &   0.55  & 0.20   \\
HD\,184499 	         & $-0.27$ & 3  & $ 1.02$  & 7 & 0.16    & $ 2.20$ &   2.23  &   2.23  &   2.22  & 0.17   \\
HD\,194598 	         & $-0.77$ & 11 & $ 0.18$  & 7 & 0.14    & $ 1.38$ &   1.46  &   1.44  &   1.40  & 0.14   \\
HD\,201891 	         & $-0.64$ & 9  & $ 0.51$  & 7 & 0.14    & $ 1.72$ &   1.79  &   1.76  &   1.73  & 0.15   \\
HD\,221377 	         & $-0.63$ & 1  & $<-0.92$ & 3 & \nodata & $ 1.41$ &   1.50  &   1.50  &   1.48  & 0.19   \\
BD\,$+3\degr 740$    & $-2.16$ & 2  & $-1.14$  & 7 & 0.13    & $<0.44$ & $<0.77$ & $<0.74$ & $<0.54$ & \nodata\\
BD\,$+23\degr 3130$  & $-1.66$ & 4  & $-1.33$  & 6 & 0.14    & $-0.18$ & $-0.02$ & $-0.12$ & $-0.16$ & 0.25   \\
BD\,$+26\degr 3578$  & $-1.73$ & 8  & $-0.77$  & 7 & 0.13    & $<0.24$ & $<0.52$ & $<0.52$ & $<0.36$ & \nodata\\
BD\,$-13\degr 3442$  & $-2.22$ & 10 & $-1.29$  & 7 & 0.12    & $<0.23$ & $<0.55$ & $<0.52$ & $<0.33$ & \nodata\\
\enddata
\tablerefs{(1) \citet{sneden1979}; (2) \citet{tomkin1992}; (3) \citet{boesgaard1993}; (4) \citet{cavallo1997};
(5) \citet{molaro1997}; (6) \citet{garcialopez1998}; (7) \citet{boesgaard1999}; (8) \citet{nissen2002};
(9) \citet{fulbright2003}; (10) \citet{akerman2004}; (11) \citet{jonsell2005}; (12) \citet{tan2009}.}
\end{deluxetable}

\begin{deluxetable}{llcc}
\tablewidth{0pt}
\tablecolumns{4}
\tablecaption{Results of linear least-squares fits to the Be and B abundance data\label{table:fit}}
\tablehead{\colhead{$x$} & \colhead{$y$} & \colhead{Slope} & \colhead{Intercept}}
\startdata
\multicolumn{4}{c}{This work}\\
$\log\epsilon\mathrm{(Be)}$                & [Fe/H] & $1.11\pm0.10$ & $1.74\pm0.16$ \\
$\log\epsilon\mathrm{(B)_{LTE}}$           & [Fe/H] & $1.17\pm0.13$ & $2.96\pm0.19$ \\
$\log\epsilon\mathrm{(B)_{NLTE,S_{H}=0}}$  & [Fe/H] & $1.07\pm0.12$ & $2.92\pm0.19$ \\
$\log\epsilon\mathrm{(Be)}$                & [O/H]  & $1.33\pm0.13$ & $1.40\pm0.14$ \\
$\log\epsilon\mathrm{(B)_{LTE}}$           & [O/H]  & $1.46\pm0.16$ & $2.63\pm0.17$ \\
$\log\epsilon\mathrm{(B)_{NLTE,S_{H}=0}}$  & [O/H]  & $1.34\pm0.15$ & $2.62\pm0.16$ \\
\tableline
\multicolumn{4}{c}{\citet{rich2009}}\\
$\log\epsilon\mathrm{(Be)}$                & [Fe/H] & $0.92\pm0.04$ & $1.41\pm0.09$ \\
$\log\epsilon\mathrm{(Be)}$                & [O/H]  & $1.21\pm0.08$ & $1.32\pm0.14$ \\
\tableline
\multicolumn{4}{c}{\citet{tan2009}}\\
$\log\epsilon\mathrm{(Be)}$                & [Fe/H] & $1.10\pm0.07$ & $1.63\pm0.10$ \\
$\log\epsilon\mathrm{(Be)}$                & [O/H]  & $1.30\pm0.08$ & $1.19\pm0.08$ \\
\tableline
\multicolumn{4}{c}{\citet{smiljanic2009}}\\
$\log\epsilon\mathrm{(Be)}$                & [Fe/H] & $1.24\pm0.07$ & $1.62\pm0.08$ \\
$\log\epsilon\mathrm{(Be)}$         & [$\alpha$/H]  & $1.36\pm0.08$ & $1.38\pm0.07$ \\
\tableline
\multicolumn{4}{c}{\citet{boesgaard1999}}\\
$\log\epsilon\mathrm{(Be)}$                & [Fe/H] & $0.96\pm0.04$ & $1.41\pm0.03$ \\
$\log\epsilon\mathrm{(Be)}$                & [O/H]  & $1.45\pm0.04$ & $1.31\pm0.04$ \\
\tableline
\multicolumn{4}{c}{\citet{duncan1997}}\\
$\log\epsilon\mathrm{(B)_{LTE}}$           & [Fe/H] & $0.96\pm0.07$ & $2.59\pm0.13$ \\
$\log\epsilon\mathrm{(B)_{LTE}}$           & [O/H]  & $1.21\pm0.13$ & $2.25\pm0.16$ \\
\enddata
\end{deluxetable}

\end{document}